\documentclass[3p,times]{elsarticle}

\usepackage{ecrc}

\usepackage{epstopdf}

\volume{00}

\firstpage{1}

\journalname{Nuclear Physics A}

\runauth{Ahmed M. Hamed (for the STAR Collaboration)}

\jid{nupha}

\jnltitlelogo{Nuclear Physics A}

\usepackage{graphicx}
\usepackage{amsmath,amssymb}

\begin{document}

\begin{frontmatter}



\title{High-$p_{_{T}}$ Direct Photon Azimuthal Correlation Measurements}

\author{Ahmed M. Hamed (for the STAR\fnref{col1} Collaboration)}
\fntext[col1] {A list of members of the STAR Collaboration and acknowledgements can be found at the end of this issue.}
\address{University of Mississippi, Oxford, USA\\Texas A$\&$M University, College Station, USA}




\begin{abstract}
The azimuthal correlations of direct photons ($\gamma_{_{dir}}$) with high transverse momentum ($p_{_{T}}$), produced at mid-rapidity ($|\eta^{\gamma_{_{dir}}}|<1$) in Au+Au collisions 
at center-of-mass energy $\sqrt{s_{_{NN}}}=200$~GeV, 
are measured and compared to those of neutral pions ($\pi^{0}$) in the same kinematic range.
The measured azimuthal elliptic anisotropy of direct photon, $v_{_{2}}^{\gamma_{_{dir}}}(p_{_{T}})$, at high $p_{_{T}}$ ($8< p_{_{T}}^{\gamma_{_{dir}}}<20$~GeV/$c$) is found to be smaller than that of $\pi^{0}$ and consistent 
with zero when using the forward detectors ($2.4 <|\eta|< 4.0$) in reconstructing the event plane.
The associated charged hadron spectra recoiled from $\gamma_{_{dir}}$ show more suppression than those recoiled from $\pi^{0}$
$(I_{_{AA}}^{\gamma_{_{dir}}-h^{\pm}} < I_{_{AA}}^{\pi^{0}-h^{\pm}})$ in the 
new measured kinematic range 
$12< p_{_{T}}^{\gamma_{_{dir}},\pi^{0}}<24$~GeV/$c$ and $3< p_{_{T}}^{assoc}<24$~GeV/$c$. 
\end{abstract}


\end{frontmatter}



\section{Introduction}
\label{intro}
A major goal of measurements at the Relativistic Heavy Ion Collider
(RHIC) is to quantify the properties of the QCD matter created in
heavy-ion collisions at high energy~\cite{STAR_white}. 
Unlike quarks and gluons, photons do not fragment into hadrons and can be directly observed 
as a final state particle. Furthermore, due to their negligible coupling to the QCD matter in contrast to hadrons,
direct photons are considered as a calibrated probe for the QCD medium.

The previous measurements at RHIC indicate unexpected finite values of azimuthal elliptic anisotropy parameter 
$\it v_{_{2}}$ of charged hadrons 
at high-$p_{_{T}}$~\cite{STAR1}. 
The measured $\it v_{_{2}}$ at high-$p_{_{T}}$ is beyond the applicability of   
hydrodynamic models, and the path-length dependence of jet quenching is the only 
proposed explanation of $\it v_{_{2}}$ at high-$p_{_{T}}$~\cite{Edward}. 
The $\it v_{_{2}}^{\gamma_{_{dir}}}$ measurement would provide a gauge for the energy loss at high-$p_{_{T}}$. 

The high-$p_{_{T}}$ $\gamma_{_{dir}}$ sample unbiased spatial 
distribution of the hard scattering vertices in the QCD medium~\cite{Wang_idea}, in contrast to hadrons which suffer 
from the geometric biases.
Therefore, a comparison between the spectra 
of the away-side charged hadrons associated with $\gamma_{_{dir}}$ vs. $\pi^{0}$ 
can provide a benchmark for the energy loss and its dependence on the path-length. Although the previous measurements have indicated
similar level and pattern of suppression for the away-side of $\gamma_{_{dir}}$ and $\pi^{0}$~\cite{STAR2}, the current work
explores a softer region in the fragmentation functions where a more significant 
difference is expected~\cite{Renk_gamma0}.
\section{Analysis and Results}
\subsection{Electromagnetic neutral clusters}
The STAR detector is well suited for measuring azimuthal angular correlations 
due to the large coverage in pseudorapidity 
and full coverage in azimuth ($\phi$). 
While the Barrel Electromagnetic
Calorimeter (BEMC)~\cite{STAR_BEMC} measures the 
electromagnetic energy with high resolution, the Barrel Shower Maximum Detector (BSMD) provides fine spatial 
resolution and enhances the rejection power for the hadrons. The Time Projection Chamber (TPC: $|\eta|<1$)~\cite{STAR_TPC} identifies 
charged particles, measures their momenta, and allows for a charged-particle veto cut with the BEMC matching.   
The Forward Time Projection Chamber (FTPC: $2.4 <|\eta|< 4.0$)~\cite{STAR_FTPC} is used to measure the charged particles' momenta and to reconstruct the event plane angle. 
Using the BEMC to select events (\textit{i.e.} ``trigger") with high-$p_{_{T}}$ $\gamma$,
the STAR experiment collected an integrated 
luminosity of 23~$p$b$^{-1}$ of p+p 
collisions in 2009 and 973~$\mu$b$^{-1}$ of Au+Au 
collisions in 2011. 
In this analysis, events having a primary vertex within $\pm 55$ cm 
of the center of the TPC along the beamline in Au+Au and $\pm 80$ cm in p+p are selected. In addition, each event must have 
at least one electromagnetic cluster  
with $E_{_{T}} > 8$~GeV for the event plane correlation analysis and $E_{_{T}} > 12$~GeV for the charged hadron correlation analysis. More than 97$\%$
of these clusters have deposited energy greater than 0.5 GeV in each layer
of the BSMD. A trigger tower is rejected if it has a track 
with $p > 3.0 $~GeV/$c$ pointing to it, which reduces the number of the electromagnetic clusters by
only $\sim 7$\%. 
\subsection{$v_{_{2}}$ of neutral particles}
The $v_{_{2}}$ is determined using the standard method~\cite{flow2}: 
\begin{equation}
v_{_{2}}(p_{_{T}}) = \langle \langle \cos 2(\phi_{p_{_{T}}}-\psi_{_{\textsc{\scriptsize{EP}}}})\rangle \rangle,
\end{equation} 
where the brackets denote statistical averaging over 
particles and events, $\phi_{p_{_{T}}}$ is the azimuthal angle of the neutral particle with certain value of 
$p_{_{T}}$, and $\psi_{_{\textsc{\scriptsize{EP}}}}$ is the azimuthal angle of the event plane.
The event plane is reconstructed from charged particles, within the detector acceptance, with $p_{_{T}} < 2 $~GeV/$c$, and determined by 
\begin{equation}
\psi_{_{\textsc{\scriptsize{EP}}}} = \frac{1}{2} \tan^{-1} (\frac{\sum_{i}\sin(2\phi_{i})}{\sum_{i}\cos(2\phi_{i})} ),
\end{equation}
where $\phi_{i}$ are the azimuthal angles of all the particles used to define the event plane. In this analysis,
the charged-track quality criteria are similar to those 
used in previous STAR analyses~\cite{flow}. The event plane is measured using two different detectors in their pseudorapidity coverage:
1) using all the selected tracks inside the TPC,
and 2) using all tracks inside the FTPC 
in order to reduce the ``non-flow" contributions (azimuthal correlations not related to the event plane). 
Since the event plane is only an approximation to the true reaction plane, 
the observed correlation is divided by the event plane resolution. The event plane resolution is estimated
using the sub-event method in which the full event is 
divided up randomly into two sub-events 
as described in~\cite{flow2}. Biases due to the non-uniform acceptance of the detector are removed according to the method in~\cite{shift}.
\subsection{Azimuthal correlations of a neutral trigger particle with charged hadrons}
The azimuthal correlations of a neutral trigger particle with charged hadrons, measured as the number
of associated particles per neutral cluster per $\Delta\phi$ (``correlation functions''), 
are used in both p+p and 
Au+Au collisions to determine the (jet) associated particle yields 
in the near- ($\Delta\phi\sim$ 0) and away-sides ($\Delta\phi\sim$ $\pi$).
The near- and away-side yields, $Y^{n}$ and $Y^{a}$, of associated particles per trigger are extracted by 
integrating the $\mathrm (1/N_{_{trig}}) dN/d(\Delta\phi)$ distributions
over $\mid\Delta\phi\mid$ $\leq$~0.63 and $\mid\Delta\phi -\pi\mid$  $\leq$~0.63, respectively. 
The yield is corrected for the tracking efficiency of charged particles as a function of event multiplicity. 
\subsection{Transverse shower profile analysis}
A crucial part of the analysis is to discriminate between showers from $\gamma_{_{dir}}$ and 
two close $\gamma$'s from high-$p_{_{T}}$ $\pi^{0}$ symmetric decays. 
At $p_{_{T}}^{\pi^0} \sim 8$~GeV/$c$, the angular separation between the two $\gamma$'s 
resulting from a $\pi^{0}$ decay is small, but a $\pi^{0}$ shower is generally broader than a single $\gamma$ shower. 
The BSMD is capable of 
$2\gamma$/$1\gamma$ separation up to $p_{_{T}}^{\pi^0} \sim 24$~GeV/$c$ due 
to its high granularity ($\Delta\eta\sim 0.007$, $\Delta\phi\sim 0.007$). 
The shower shape is quantified as the cluster energy, 
measured by the BEMC, normalized by the position-weighted energy moment, 
measured by the BSMD strips~\cite{STAR2}.
The shower profile cuts were tuned to obtain a nearly $\gamma_{_{dir}}$-free 
($\pi^{0}_{_{rich}}$) sample and a sample rich in $\gamma_{_{dir}}$ ($\gamma_{_{rich}}$). 
Since the shower-shape analysis 
is only effective for rejecting two close $\gamma$ showers, the $\gamma_{_{rich}}$ sample 
contains a mixture of direct photons and contamination from 
fragmentation photons ($\gamma_{_{frag}}$) and photons from asymmetric hadron ($\pi^0$ and $\eta$) decays.

The $v_{_{2}}^{\gamma_{_{rich}}}$ and $v_{_{2}}^{\pi^{0}}$ are measured as discussed in section 2.2 and  
the away (near)-side yields of associated particles 
per $\gamma_{_{rich}}$ and $\pi^{0}_{_{rich}}$ triggers ($Y^{a(n)}_{\gamma_{_{rich}}+h}$ and $Y^{a(n)}_{\pi^{0}_{_{rich}}+h}$) are measured as discussed in section 2.3.
\subsection{\it v$_{_{2}}$ of direct photons}
Assuming zero near-side yield for $\gamma_{_{dir}}$ triggers and a  
sample of $\pi^{0}_{_{rich}}$ free of $\gamma_{_{dir}}$, the $\it v_{_{2}}^{\gamma_{_{dir}}}$ is given by:
\begin{equation}
v_{_{2}}^{\gamma_{_{dir}}}=\frac {v_{_{2}}^{\gamma_{_{rich}}}- {\cal{R}}v_{_{2}}^{\pi^{0}_{_{rich}}}} {1-\cal{R}}, 
\end{equation}
where $\cal{R}$=$\frac{N^{\pi^{0}_{_{rich}}}}{N^{\gamma_{_{rich}}}}$, and the numbers of
$\pi^{0}_{rich}$ and $\gamma_{_{rich}}$ triggers
are represented by $N^{\pi^{0}_{_{rich}}}$ and 
$N^{\gamma_{_{rich}}}$ respectively. 
Although the $\cal{R}$ quantity approximates all background triggers in the $\gamma_{_{rich}}$ sample to the measured 
$\pi^{0}_{_{rich}}$ triggers, all background to $\gamma_{_{dir}}$ is subtracted assuming that 
all background triggers have the same correlation function
as the $\pi^{0}_{_{rich}}$ sample~\cite{STAR2}.
The value of $\cal{R}$ is measured in~\cite{STAR2} and 
found to be 
$\sim 30\%$ in central Au+Au. 
In Eq. 3 all background sources 
for $\gamma_{_{dir}}$ are assumed to have the 
same $\it v_{_{2}}$ as the measured $\pi^{0}$.\\ 
\\

\begin{figure}[ht]
\begin{minipage}[b]{0.32\linewidth}
\centering
\includegraphics[width=1.15\textwidth,height=5cm, angle =-90]{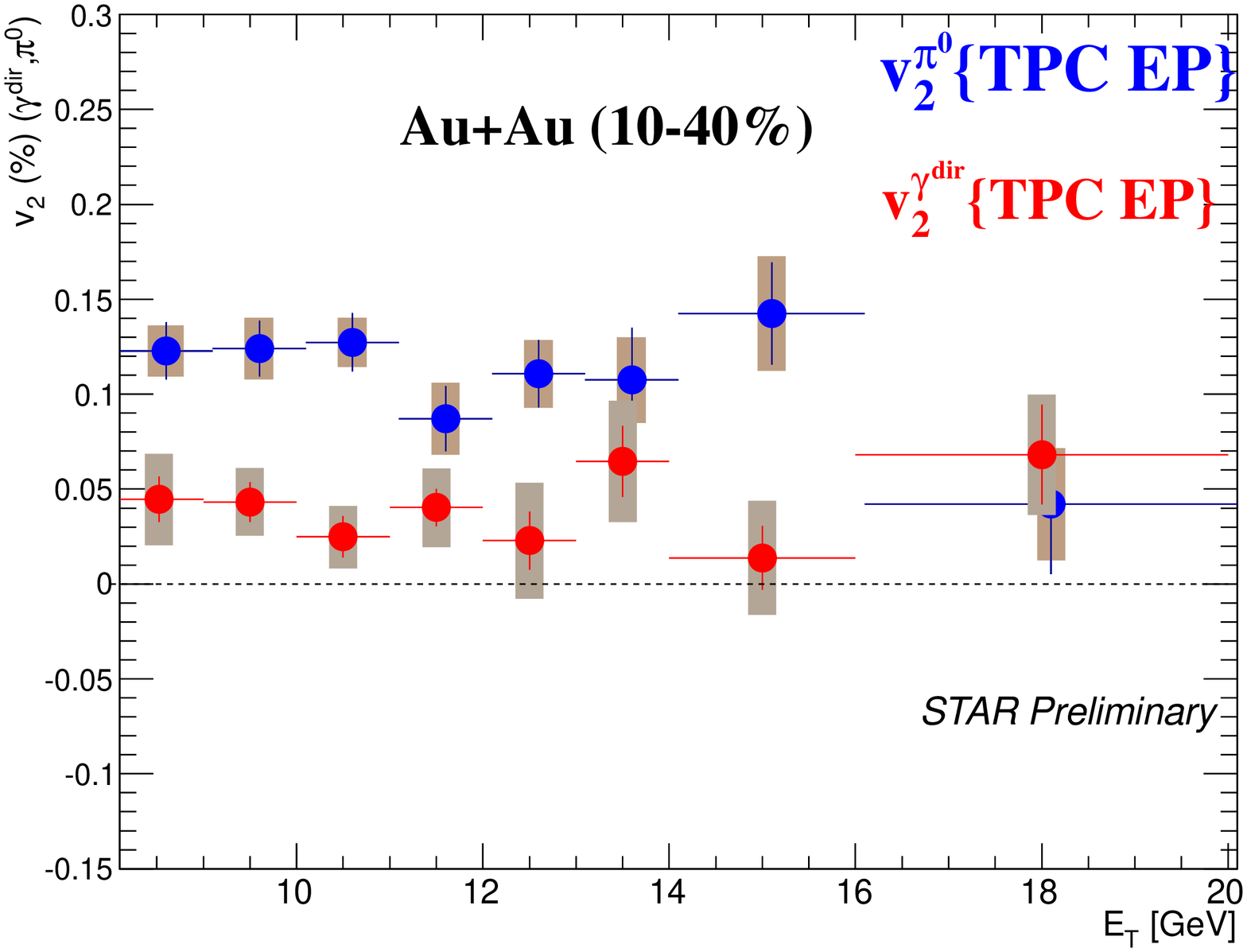}
\caption{$v_{_{2}}$ of
	$\pi^{0}$, and $\gamma_{_{dir}}$ in 10-40$\%$ centrality of Au+Au collisions at $\sqrt{s_{_{NN}}}=200$~GeV using TPC. Boxes show the systematic errors.}
\end{minipage}
\hspace{0.15cm}
\begin{minipage}[b]{0.32\linewidth}
\centering
\includegraphics[width=1.15\textwidth,height=5cm, angle =-90]{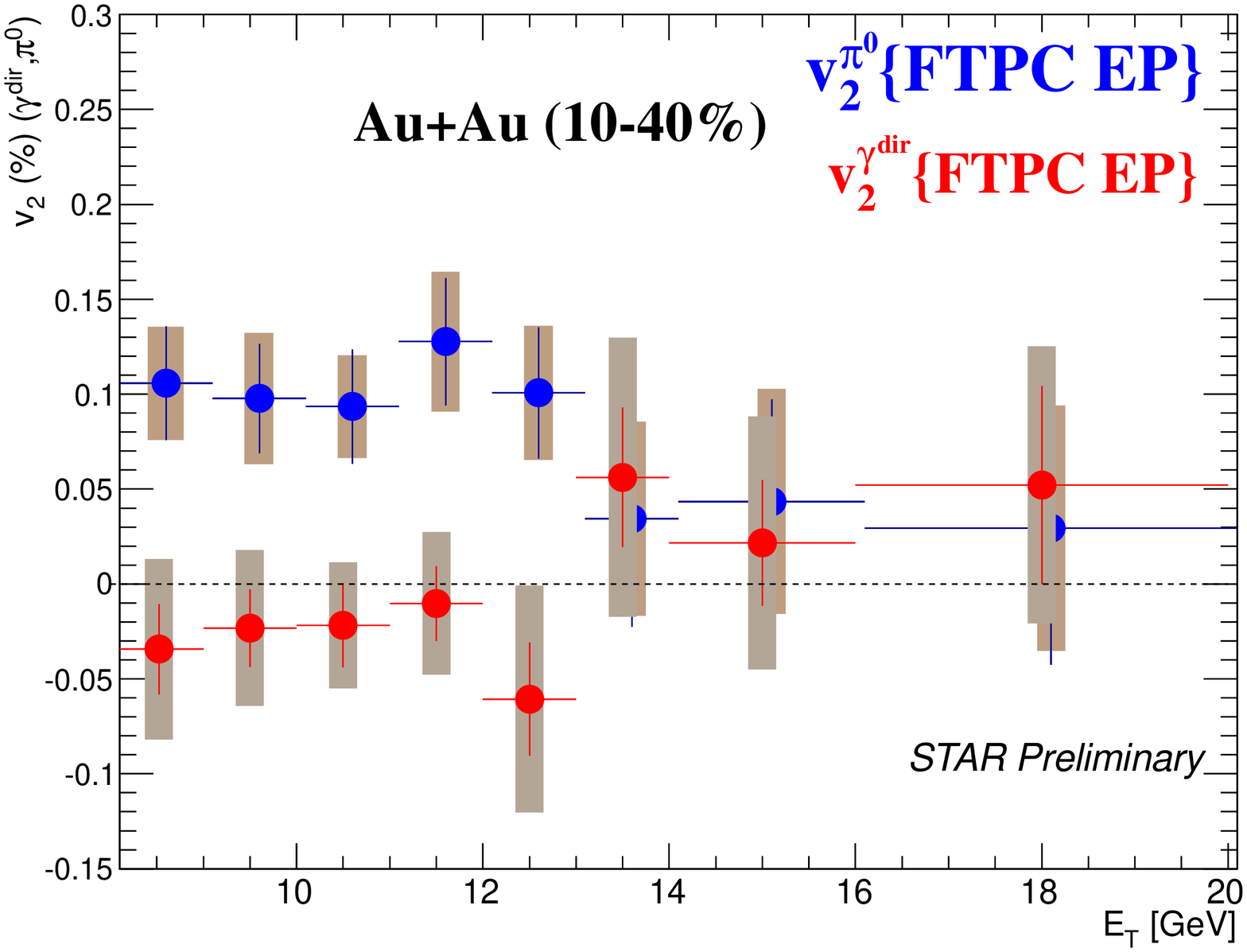}
\caption{$v_{_{2}}$ of
	$\pi^{0}$, and $\gamma_{_{dir}}$ in 10-40$\%$ centrality of Au+Au collisions at $\sqrt{s_{_{NN}}}=200$~GeV using FTPC. Boxes show the systematic errors.}
\end{minipage}
\hspace{0.15cm}
\begin{minipage}[b]{0.32\linewidth}
\centering
\includegraphics[width=1.15\textwidth,height=5cm, angle =-90]{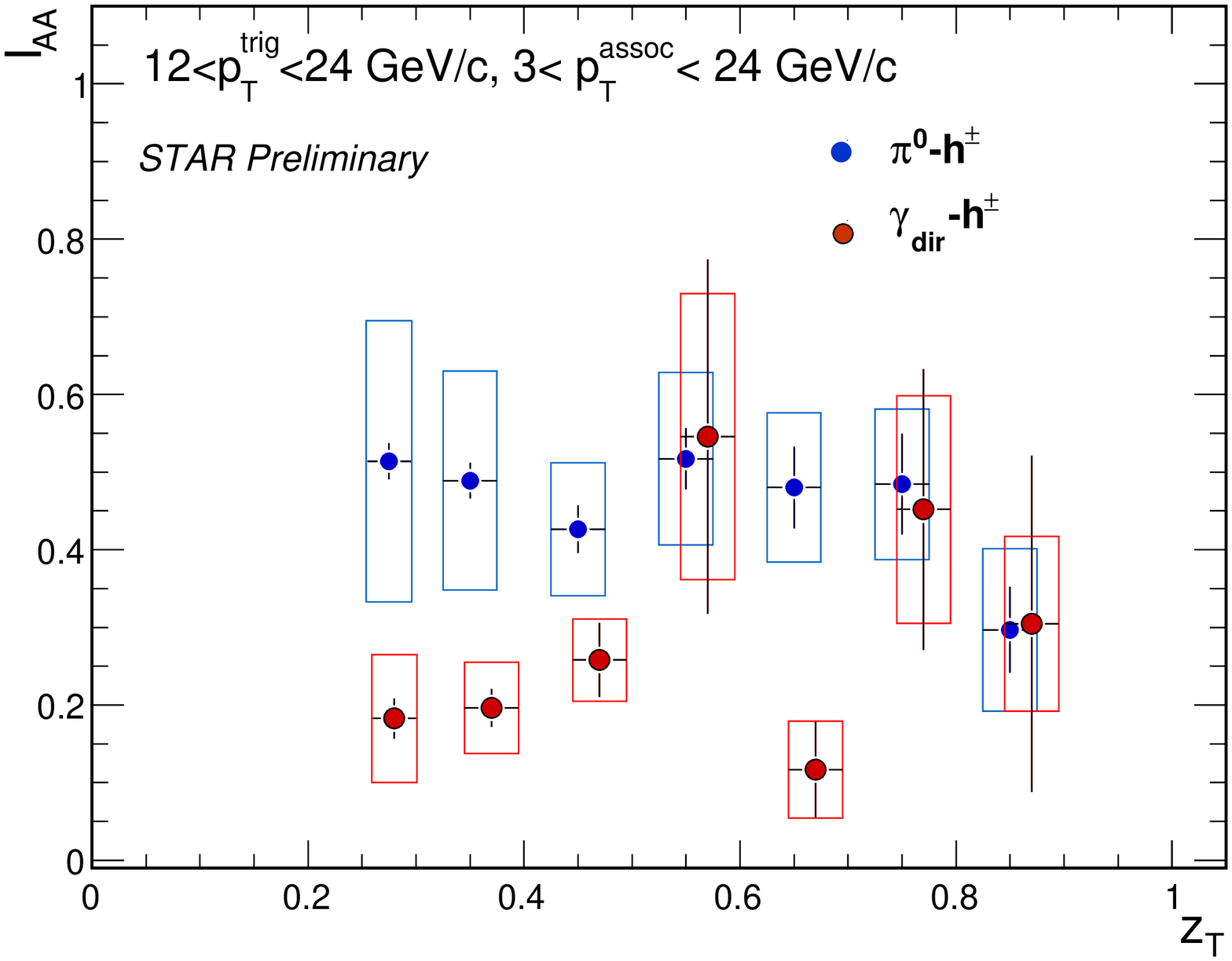}
\caption{The $z_{_{T}}$ dependence of $I_{_{AA}}$ for $\gamma_{_{dir}}$ and $\pi^{0}$ triggers in Au+Au 0-10$\%$ centrality. Boxes show the systematic errors.}
\end{minipage}
\end{figure}
Figures 1 and 2 show the $\it v_{_{2}}^{\pi^{0}}$ and $v_{_{2}}^{\gamma_{_{dir}}}$ for ($8< p_{_{T}}^{\gamma_{_{dir}}}<20$~GeV/$c$) 
from Au+Au data set of Run 2011 using the TPC ($|\eta|<1$) and FTPC ($2.4 <|\eta|< 4.0$). 
The 
results of Fig. 1 are consistent with those from different STAR data set of Run 2007~\cite{ahmed}, and the results of Fig. 2 agree with other measurements~\cite{ALICE,PHENIX}. 
While using the FTPC in determining the event plane (Fig. 2) the $\it v_{_{2}}^{\gamma_{_{dir}}}$ is consistent with zero. Assuming the dominant source of direct photons is
prompt hard production, the zero value implies no remaining bias in the event-plane determination. Accordingly, the measured value of 
$\it v_{_{2}}^{\pi^{0}}$ would be the effect of path-length dependent energy loss. 
\subsection{Extraction of $\gamma_{_{dir}}$ associated yields \label{subsec:gammadir}}
Assuming zero near-side yield for $\gamma_{_{dir}}$ triggers and a  
sample of $\pi^{0}_{_{rich}}$ free of $\gamma_{_{dir}}$, 
the away-side yield of hadrons correlated with the $\gamma_{_{dir}}$ is 
extracted as
\begin{align}
Y_{\gamma_{_{dir}}+h}&=\frac{Y^{a}_{\gamma_{_{rich}}+h}-{\cal{R}} Y^{a}_{\pi^{0}_{_{rich}}+h}}{1-{\cal{R}}}, \nonumber \\ 
{\rm where~}
{\cal{R}}&=\frac{N^{\pi^{0}_{_{rich}}}}{N^{\gamma_{_{rich}}}}=\frac{Y^{n}_{\gamma_{_{rich}}+h}}{Y^{n}_{\pi^{0}_{_{rich}}+h}},
{\rm ~~~and~} 1-{\cal{R}}=\frac{N^{\gamma_{_{dir}}}}{N^{\gamma_{_{rich}}}}. 
\label{eq:gammadir}
\end{align}
Here, $Y^{a(n)}_{\gamma_{_{rich}}+h}$ and $Y^{a(n)}_{\pi^{0}_{_{rich}}+h}$ are 
the away (near)-side yields of associated particles 
per $\gamma_{_{rich}}$ and $\pi^{0}_{_{rich}}$ triggers, respectively.  The 
ratio ${\cal{R}}$ is equivalent to the fraction of ``background'' triggers 
in the $\gamma_{_{rich}}$ trigger sample, and 
$N^{\gamma_{_{dir}}}$ and $N^{\gamma_{_{rich}}}$ are the numbers of $\gamma_{_{dir}}$ and $\gamma_{_{rich}}$ triggers, respectively. The value of ${\cal{R}}$
is found to be $\sim 55\%$ in p+p and decreases to $\sim 30\%$ in central Au+Au with little dependence on $p_{T}^{trig}$. 
All background to $\gamma_{_{dir}}$ is subtracted with the assumption that 
the background triggers have the same correlation function
as the $\pi^{0}_{_{rich}}$ sample.

In order to quantify the away-side suppression, 
we calculate the quantity $I_{_{AA}}$, which is defined as the ratio of the integrated yield of the away-side associated 
particles per trigger particle in Au+Au to that in p+p collisions. 
The values of $I_{_{AA}}^{\gamma_{_{dir}}-h^{\pm}}$ and $I_{_{AA}}^{\pi^{0}-h^{\pm}}$, as shown in Fig. 3,  are $z_{_{T}}$ ($z_{_{T}} = p_{_{T}}^{assoc}/p_{_{T}}^{trig})$ independent in agreement with results of~\cite{STAR2} where the 
recoiled parton from $\gamma_{_{dir}}$ and $\pi^{0}$ experience constant fractional energy loss in the QCD medium. It is also observed that 
the charged hadron spectra recoiled from $\gamma_{_{dir}}$ show unexpectedly more suppression than those recoiled from $\pi^{0}$ $(I_{_{AA}}^{\gamma_{_{dir}}-h^{\pm}} < I_{_{AA}}^{\pi^{0}-h^{\pm}})$ within the covered 
kinematics range 
$12< p_{_{T}}^{\gamma_{_{dir}},\pi^{0}}<24$~GeV/$c$ and $3< p_{_{T}}^{assoc}<24$~GeV/$c$.
\section{Conclusions}
The STAR experiment has reported the first $\it v_{_{2}}^{\gamma_{_{dir}}}$ at high-$p_{_{T}}$ ($8< p_{_{T}}^{\gamma_{_{dir}}}<20$~GeV/$c$), and explored new kinematic range 
($12< p_{_{T}}^{\gamma_{_{dir}},\pi^{0}}<24$~GeV/$c$) and ($3< p_{_{T}}^{assoc}<24$~GeV/$c$) for $I_{_{AA}}$ measurements of $\gamma_{_{dir}}-h$ correlations at $\sqrt{s_{_{NN}}}=200$~GeV. 
Using the mid-rapidity detectors in determining the event plane, the measured value of $\it v_{_{2}}^{\gamma_{_{dir}}}$ is non-zero, and is probably due to biases in the event-plane determination. 
Using the forward detectors in determining the event plane could eliminate remaining biases, and the measured $\it v_{_{2}}^{\gamma_{_{dir}}}$ is consistent with zero.
The zero value of 
$\it v_{_{2}}^{\gamma_{_{dir}}}$ suggests a negligible contribution of jet-medium photons~\cite{Theory2}, and negligible effects 
of ${\gamma_{_{frag}}}$~\cite{Theory1} on the $v_{_{2}}^{\gamma_{_{dir}}}$ over the covered kinematics range. The measured finite value of 
$v_{_{2}}^{\pi^{0}}$, using the forward detectors in determining the event plane, is apparently due to the path-length dependence of energy loss.
The $\gamma_{_{dir}}-h$ correlation 
results indicate that the associated charged hadron spectra recoiled from $\gamma_{_{dir}}$ show more suppression than those recoiled from 
$\pi^{0}$ $(I_{_{AA}}^{\gamma_{_{dir}}-h^{\pm}} < I_{_{AA}}^{\pi^{0}-h^{\pm}})$ within the covered kinematic range, in contrast to the theoretical predictions~\cite{Wang}.
The disagreement with the theoretical expectations may indicate that the lost energy is distributed to lower $p_{_{T}}$ of the associated particles in the case of a $\gamma_{_{dir}}$ trigger than a $\pi^{0}$ trigger. 
To further test this, one must explore the region of low $p_{_{T}}^{assoc}$ and $z_{_{T}}$.

\end{document}